\documentstyle[prl,aps]{revtex}
\tighten
\begin{document}
\draft

\title{ 
Current drag in capacitevly coupled Luttinger constrictions
       }
\author{ Yuli V. Nazarov}
\address{
Department of applied Physics and Delft Institute of
Micro-electronics and Submicron-technology (DIMES), 
Delft University of Technology,
 2628 CJ Delft, the Netherlands
	}
\author{ D. V. Averin}
\address{Department of Physics, State University of New York, 
Stony Brook, NY 11794, USA}
\maketitle
\begin{abstract}
We study the current drag in the system of two electrostatically
coupled finite 1D electron channels. We present the perturbation 
theory results along with the results for two non-perturbative 
regimes. It is shown that the drag may become absolute, that is, 
the currents in the channels are equal in a finite window of the bias 
voltages. 
\end{abstract} 

\pacs{71.10.Pm, 73.23.-b, 73.23.Ad }

The recent experiments with two capacitevly coupled 2DEG
\cite{Twodeg} demonstrated that, owing to Coulomb interaction,
electrons moving in one of the 2DEG drag the electrons of
the adjacent 2DEG. Therefore the system work as a d.c. current
transformer. The transformation coefficient was however fairly 
small. Later it has been shown that in a specially designed 
coupled 1D arrays of ultrasmall tunnel junctions Coulomb 
interaction may lead to the absolute current drag \cite{We}. This 
means that the electric currents in two capacitevily coupled 
circuits are equal in magnitude in a certain region of voltages 
applied to the circuits, and the system may work as a current 
copier. This prediction has been very recently confirmed 
\cite{Marco} experimentally. 

The mechanism of the absolute current drag in small tunnel 
junctions is however quite different from the momentum transfer 
physics that is responsible for the current drag in the 2DEG 
systems. The aim of the present work is to show that in the 
1D electron channels, the momentum-transfer mechanism can also lead 
to the absolute current drag. We consider a system of two channels of 
finite length $L$ that are adiabatically connected to the reservoirs 
of effectively non-interacting electrons (Fig.1). The channels 
are assumed to be coupled only by the Coulomb interaction. 
Further, we assume no impurities in the channels that would cause 
electron backscattering. Being inspired by recent advances 
\cite{Tarucha} in fabrication of relatively long one-channel 
quantum wires, we hope that such systems can be successfully 
fabricated in the nearest future.

To get an intuitive feeling of how the absolute current drag may 
occur in such a system, let us consider the case of strong electron 
repulsion. In this case, electrons in each channel form a rigid 
Wigner lattice. Provided the channels are close to each other, the 
repulsion coordinates not only the positions of electrons inside the 
same channel but also the positions of neighboring electrons in the 
other channel. (Fig. 1) Now, if the electrons in one of the channels move, 
electrons in the other channel must follow their motion, so that 
the electric currents in the channels are equal. (Fig. 1)

Such a simple model says very little about realistic 1D channels
where electrons are subjected to strong quantum fluctuations.  
Below we develop a consistent theory of current drag  
which accounts for these fluctuations. The main prediction of the 
theory is that the (almost) absolute current drag survives quantum 
fluctuations. It can occur even if the repulsive interaction is weak 
provided the channels are sufficiently long.

We model each of the two coupled 1D conductors as ``Luttinger 
constriction'' with spinless electrons, i.e., use the standard 
Hamiltonian \cite{Lutt}: 
\begin{equation} 
H= \int \frac{dx}{2\pi}  u \left[ g\Pi^2+\frac{(\nabla \Theta)^2}{g}
\right] \, .
\label{10} \end{equation}  
Here $\Theta(x)$ is the displacement of electrons in the conductor 
normalized in such a way that the local fluctuation of electron 
density is $\delta n= -\nabla \Theta /\pi$, and $\Pi(x)$ is the 
conjugated momentum density: 
$[\Theta(x),\Pi(x')]= i\pi \hbar \delta(x-x')$.  
The velocity $u$ of excitations in the conductors depends on $x$: 
in the constriction region of length $L$ the electron-electron 
interaction gives an essential contribution to compressibility and 
$u$ is larger than outside of the constriction, where interaction is 
screened out and $u$ is equal to the Fermi velocity $u_F$. 
Parameter $g$ in the Hamiltonian (\ref{10}) characterizes the 
strength of the interaction, $g=u_F/u$.  

The relevant part of the interaction of the two conductors 
responsible for the momentum transfer between them can be written as 
\begin{equation} 
H_c =J\int_{-L/2}^{L/2} \frac{dx}{2\pi} \cos [2(k_{F1}-k_{F2})x+
2(\Theta_1(x) - \Theta_2(x)) ] \, 
\label{11} \end{equation} 
where subscripts $1,2$ refer to the two conductors, $k_{Fj}=\pi n_j$ 
are the Fermi wavevector, and $J$ is the interaction constant, precise 
value of which depends on the specific electrostatic configuration 
of the system. In particular, it depends on the distance $d$ between 
the two channels, decreasing roughly as $\exp(-dn)$ at large $d$, 
where $n$ is the average electron concentration in the channels.    

When the coupling is weak, electrons in the two channel move  
independently and can carry generically different currents $I_1$ and 
$I_2$. When the currents are different however, momentum transfer 
between the channels leads to the appearance of an extra voltage 
drop $\delta V_j$ across each channel, in addition to the ``ideal''  
Landauer value $hI_j/e^2$. The extra voltages $\delta V_j$ can be 
found from the difference of chemical potentials at the ends of the 
channels: 
\begin{equation}
\delta V_j=\frac{\kappa_j}{e} \langle \delta n_j(-L/2)-
\delta n_j(L/2) \rangle \, .
\label{13} \end{equation} 
where $\kappa$'s are compressibilities of electron liquids in the  
channels. Writing the equations of motion for phases $\Theta$ that 
follow from the Hamiltonian $H=H_1+H_2 +H_c$: 
\begin{equation} 
\ddot{\Theta}_j-[u_j^2 \Theta_j']'= (-1)^j J u_F \sin 
[2(k_{F1}-k_{F2})x+2(\Theta_1(x) - \Theta_2(x)) ] \, , 
\label{133} \end{equation} 
and using the fact that in the stationary case $\langle \ddot{\Theta} 
\rangle=0$, we get that $\delta V_2=-\delta V_1 \equiv V$, and: 
\begin{equation} 
V =\int_{-L/2}^{L/2} dx \frac{F(x)}{e} \, , \;\;\;\;\; F(x) = {J} 
\langle \sin (2(k_{F1}-k_{F2})x+2(\Theta_1(x) - \Theta_2(x)) \rangle \, .  
\label{14} \end{equation}  

In the first order of the perturbation theory in $H_c$ we get: 
\begin{equation} 
F(x) = \frac{J^2}{2}\mbox{Re} \int_{-L/2}^{L/2}  dx' \int_0^{\infty} 
dt e^{i(k(x-x')+\Omega t)} e^A(e^B-e^{-B}) \, ,  
\label{15} \end{equation}  
where $k=2(k_{F1}-k_{F2})$, and $ A(x,x',t) =  \langle 
[\varphi,\varphi']_+ - \varphi^2 - \varphi'^2 \rangle /2$, 
$B(x,x',t) = \langle [\varphi,\varphi'] \rangle /2$ .
Here $\varphi =\varphi(x,t)$, $\varphi' =\varphi(x',0)$, and 
the phase $\varphi(x,t)$ is introduced by the relation: 
\begin{equation} 
2(\Theta_1(x,t) - \Theta_2(x,t)) = \Omega t +\varphi(x,t) \, ,  
\;\;\;\; \Omega =2\pi (I_1-I_2)/e \, . 
\label{16} \end{equation}  
The average $\langle ... \rangle$ in the definitions of the 
correlators $A$ and $B$ is taken over the equilibrium fluctuations of 
$\varphi$. The assumption of equilibrium is legitimate since after we 
separated the linearly growing part $\Omega t$ in eq.\ (\ref{16}) the 
dynamics of $\Theta$'s in the zero order in $H_c$ does not have any 
perturbations that would drive the system out of the equilibrium. 

The equilibrium correlators $A,B$ can be expressed in a standard way 
in terms of the spectral densities $\rho_j(x,x',\omega)$:    
\begin{equation} 
A(x,x',t)\pm B(x,x',t)= 2 \sum_{j=1,2} \int d\omega \left[ 
e^{i\omega t} \rho_j (x,x',\omega)(\coth\frac{\omega}{2T} \pm 1) 
- \rho_j (x,x,\omega) -\rho_j (x',x',\omega) \right] \, , 
\label{17} \end{equation} 
where spectral densities are $\rho_j = 
(1/\pi)\mbox{Im} G_R^{(j)}(x,x',\omega)$. The retarded Green's 
functions $G_R^{(j)}$ of the phases $\Theta_j$ satisfy the equations 
(\ref{133}) with $J=0$, radiative boundary conditions at $x\rightarrow 
\pm \infty$, and delta-functional source term in the right-hand-side.  
Such equations can be solved explicitly for our model 
of piecewise constant $u(x)$ and give for $x,x'$ inside the 
constriction: 
\begin{equation} 
\rho (x,x',\omega) = \frac{g^2}{\omega} \frac{(1+g^2)\cos\nu(x-x') 
+(1-g^2)\cos\nu L \cos\nu(x+x')}{(1+g^2)^2 \sin^2\nu L + 4g^2 
\cos^2 \nu L}\, , \;\;\;\; \nu=\omega/u \, .  
\label{18} \end{equation} 
 
Equation (\ref{14}) combined with eqs.\ (\ref{17}) and (\ref{18}) 
allows us to find the induced voltage $V$ as a function of the 
current difference $I=I_1-I_2$ in various regimes. We consider 
first identical channels at zero temperature. In the non-interacting 
case $g=1$ we get then from eq.\ (\ref{14}): 
\begin{equation} 
V= \frac{\pi J^2 L u \Omega^2}{2e E_F^4} \int_0^{L \Omega/u} 
\frac{dx}{x} \frac{\sin x -x\cos x}{x^2}(1-\frac{x u}{L \Omega}) \, , 
\;\;\;\;\; \, ,   
\label{19} \end{equation} 
where $E_F$ is the cutoff frequency of the excitation 
spectrum of the conductors, being of the order of Fermi energy. 
Equation (\ref{19}) shows that at small current differences $\Omega$, 
for which $\Omega \ll u/L$, the voltage $V$ grows as $\Omega^3$, and 
this growth slows down to $\Omega^2$ at large $\Omega$. 

At small frequencies the spectral density (\ref{18}) is dominated 
by the leads and approaches its non-interacting value $1/2\omega$ 
at arbitrary interaction strength inside the constriction. Since at 
small current differences $\Omega$ the system does not have enough energy 
to excite high-frequency modes in the constrictions this means that,
the dependence $V\propto \Omega^3$ is valid at sufficiently small $\Omega$ 
for any $g$. However the range of the validity of this dependence 
shrinks with growing interaction strength (decreasing $g$) to 
$\Omega \ll gu/L$. 
 
To see how $V$ behaves at large current differences ($\Omega \gg u/L$) 
we can average the rapid oscillations of spectral density (\ref{18}) 
with the period $u/L$ and also neglect the terms that oscillate rapidly 
with $x,x'$. The spectral densities reduce then to $\rho_j (x,x',\omega) = 
g_j \cos\nu(x-x') /2\omega$, and we get for $V$: 
\begin{equation} 
V=FL= \frac{J^2 L u \Gamma^2 (1-2g) \sin^2 2\pi g}{E_F^2} 
\left( \frac{\sqrt{\Omega^2-u^2k^2}}{2 E_F} \right)^{4g-2} 
\Theta(\Omega^2-u^2k^2)\, , \;\;\; g=\frac{g_1+g_2}{2} \, . 
\label{20} \end{equation} 
We see that the interaction inside the constriction changes drastically 
the behavior of $V$ as a function of $\Omega$ giving rise to power-like 
dependence. The voltage $V$ decreases at large $I$ for $g<1/2$, so that 
there is optimum $I\simeq eu/L$ for which the induced voltage is maximum. 
If we take oscillating terms into account, the decrease at large current 
differences is non-monotonic, the voltage oscillates with $\Omega$ with 
the period $u/L$. This resembles density of states behavior in finite 
Luttinger constriction \cite{Density}. All these features can be seen in 
Fig.\ 2 which presents the results of numerical calculations of $V$ as a 
function  of $I_1-I_2$ from Eqs.\ 6-9. 


Another important feature of Eq.\ \ref{20} is that the voltage $V$ grows 
with increasing $L$ at any given current difference $\Omega$ regardless 
of $g$. This suggests that Eq.\ \ref{20} is valid only for sufficiently 
short channels, when the voltage $V$ is small perturbation. In longer 
channels the distribution of the voltage along the channel and  
associated with it difference of electron densities in the two channels 
should be determined self-consistently. The two quantities are related 
through the differential version of Eq.\ \ref{13}, which for identical 
channels has the form $V'(x)=-\kappa k'(x)/2\pi e$. The variation of the 
density difference can be described by the steady-state form of Eq. 
\ref{133} which can be written as follows: 
\begin{equation}
k'(x) = \frac{4g}{u} F(k(x),\Omega) \, . 
\label{Fric} \end{equation}   
Here the force $F$ is given by Eq. \ref{20} but now with $local$ $k(x)$. 
In simple terms, Eq.\ \ref{Fric} describes how the friction between the 
moving liquids compresses one of them and stretches another one, and how 
this compression/stretch affects the magnitude of the friction itself.  

Equation \ref{Fric} with the force given by eq.\ \ref{20} shows that 
in long channels, $L \gg L_{\Omega}$, where $L_{\Omega}^{-1} \equiv 16 
g J^2 u (\Omega/2 E_F)^{4g} /\Omega^3$ is characteristic length of 
variation of $k(x)$, the density difference approaches 
the value $\Omega/\pi u$ and saturates. This means that the extra voltage 
difference drops at distances of the order of $L_{\Omega}$ near the edge 
of the interacting region and is equal to $h (I_1-I_2)/e^2 2 g$. The total 
voltage drops across the channels are given then by ``quasi Landauer'' 
relations
\begin{equation}
e^2 V_1/h = I_1 +\frac{1}{2g} (I_1-I_2) \, , \;\;\;\;\;\; 
e^2 V_2/h = I_2 +\frac{1}{2g} (I_2-I_1) \, .  
\end{equation}
We see from this equations that the transconductance of the two channels 
depends strongly on the interaction strength $g$ inside the channels, and 
for strong interaction, $g\rightarrow 0$, the currents in the two channels 
are almost equal in the wide range of bias voltages.  

Another nonperturbative regime in which the currents in the two channels 
are almost precisely equal is the limit of strong coupling of the two short 
channels with $L\ll L_Q$. This regime corresponds to the classical limit of 
the sine-Gordon model.\cite{Sg} In the limit of $g \rightarrow 0$, one
can treat $\varphi$ as (almost) classical variable that takes on  
$2\pi \times$integer values. In our original model, this means that the
Luttinger liquids, or electron chains, are strictly correlated. However,
they can move together and carry {\it equal} currents in both 
constrictions. The relation between current and voltages in this case 
is readily obtained from the fact that, as follows from Eq. (\ref{13}), 
the deviations from Landauer voltages are opposite in both channels, 
$\delta V_1=-\delta V_2$. From this we obtain $I_1=I_2= e^2 (V_1+V_2)/2h$. 

This is, of course, an ideal situation, and at finite $g$ and $L$ the 
currents are not precisely equal. Let us show that, in the wide region of 
parameters, the current difference is {\it exponentially small}. To this 
end, we switch from the Hamiltonian to Lagrangian in imaginary time. This 
representation is standard for calculations of rates of classically 
forbidden processes. The Lagrangian describing the relative motion of the 
two electron liquids within the interaction region takes the standard 
sine-Gordon form,
\begin{equation}
S [ \varphi(x,t)] =\int \frac{dx dt}{2\pi} \left[ 
\frac{1}{8 g} ( u {\varphi'}^2 + \frac{\dot{\varphi}^2}{u}) 
+ J \cos \varphi \right] .     
\end{equation}
Equation \ref{16} shows that the shift of $\phi$ at the boundary by $4\pi$ 
corresponds to transfer of one electron into the channel 1 accompanied by 
the transfer of an electron out of the channel 2. This suggests that,
in order to describe the interaction of the electron channels with the 
electrodes, we should add to the Lagrangian the boundary term
\begin{equation}
S_b = \int \frac{dt}{4 \pi} \left[ (\mu_2^R-\mu_1^R) \varphi(-L/2)
+ (\mu_1^L-\mu_2^L) \varphi(L/2) \right] \, ,
\end{equation}
where $\mu_{1,2}^{L,R}$ are the chemical potentials of the left/right 
electrode of the channel 1 and 2. 

What are the processes that spoil the ideal current copying? Obviously,
those are slips of $\varphi$ by $2\pi$. If the voltages applied to the
constriction are sufficiently low, the slip should happen at once
throughout the interacting region. Otherwise a soliton would have been
created within the interacting region as a result of the slip. 
This would cost energy $E_s=\sqrt{Ju/\pi g}$ and is forbidden by the  
energy considerations. The sudden slip throughout the interacting region 
may be viewed as a virtual soliton transfer via the "barrier" $E_s$ 
provided the barrier length exceeds the soliton size $\sqrt{u/Jg}$. 
The rate of such a process is determined by the saddle-point configuration 
of $\phi(x,\tau)$ as sketched in Fig.\ 3. The action is contributed by 
two soliton "walls" and  the boundary term ($\mu = \max( (\mu_1^L-\mu_2^L)/2, 
(\mu_1^R-\mu_2^R)/2 )$),
\begin{equation}
S= 2 E_s \sqrt{\tau^2 +(L/u)^2} - 2 \tau \mu \, .
\end{equation}
Minimization of this expression with respect to $\tau$ yields
\begin{equation}
S_0 = 2 \sqrt{E_s^2 -\mu^2} L /u \, ,
\label{Snul} \end{equation}
and $I_1 -I_2 \propto \exp(-S_0)$. We see that the current difference
is suppressed exponentially provided $L \gg u/E_s$ and $\mu_{L,R} < E_s$.
The common wisdom of the sine-Gordon model suggests that these results 
are in fact valid not only in the classical limit $g \rightarrow 0$ but 
also at any $g<1$, provided $E_s$ is properly renormalized corresponding 
to the actual soliton energy $E_s(g)$. The concrete formula for $E_s(g)$ 
reads
\begin{equation}
	E_s(g) \propto E_s {(E_s/E_F)}^{g/1-g} 
\end{equation}
with the coefficient depending upon cutoff details. There is an interesting 
analogy between the soliton and Dirac particle with the dispersion relation 
$E(k)=\sqrt{E_s^2+u^2 k^2}$ in the interacting region and $E(k)=uk$ beyond 
that. At energies $E<E_s$ the particle would tunnel through the interacting 
region. The energy- and $L$-dependence of such a tunneling process coincide 
exactly with that of Eq.\ \ref{Snul}.

In conclusion, we have considered current drag in the system of 
two capacitevely coupled Luttinger constrictions. We have presented
perturbative results and have found two nonperturbative regimes.
In one of the regimes, the system works as an ideal current copier of 
exponential accuracy.
 
We wish to thank A.A. Odintsov for his contribution at preliminary 
stages of this work. We are indebted to S. Tarucha for his communications 
that gave an impetus to the present research, G.E.W. Bauer, M.P.A. Fisher 
for very instructive discussions of the results. This work has been made 
possible by financial support of North Atlantic Treaty
Organization, grant No. 950279.

\begin{figure}
\caption{
The system under consideration. Two 1D channels are open to reservoirs and
separated by inslulating barrier. The electrostatic repulsion coordinates
the positions of electrons (black circles) in the channels, that may lead
to the absolute current drag.
}
\label{fig1}
\end{figure}

\begin{figure}
\caption{The perturbative regime. The induced voltage difference
is normalized by $V_0(g) \equiv J^2 u L E_F^{-2} (u/E_F L)^{4g-2}$.The curves a,b,c,d,e correspond to
the values of $g= 0.1, 0.3, 0.5, 0.6, 1.0$ respectively.}
\label{fig2}
\end{figure}

\begin{figure}
\caption{
Soliton tunneling via the interacting region. 
The optimal configuration of $\varphi(x,\tau)$ correspond to
three regions of almost constant $\varphi$ separated by V-shaped soliton "wall" (thick  line). 
}
\label{fig3}
\end{figure}

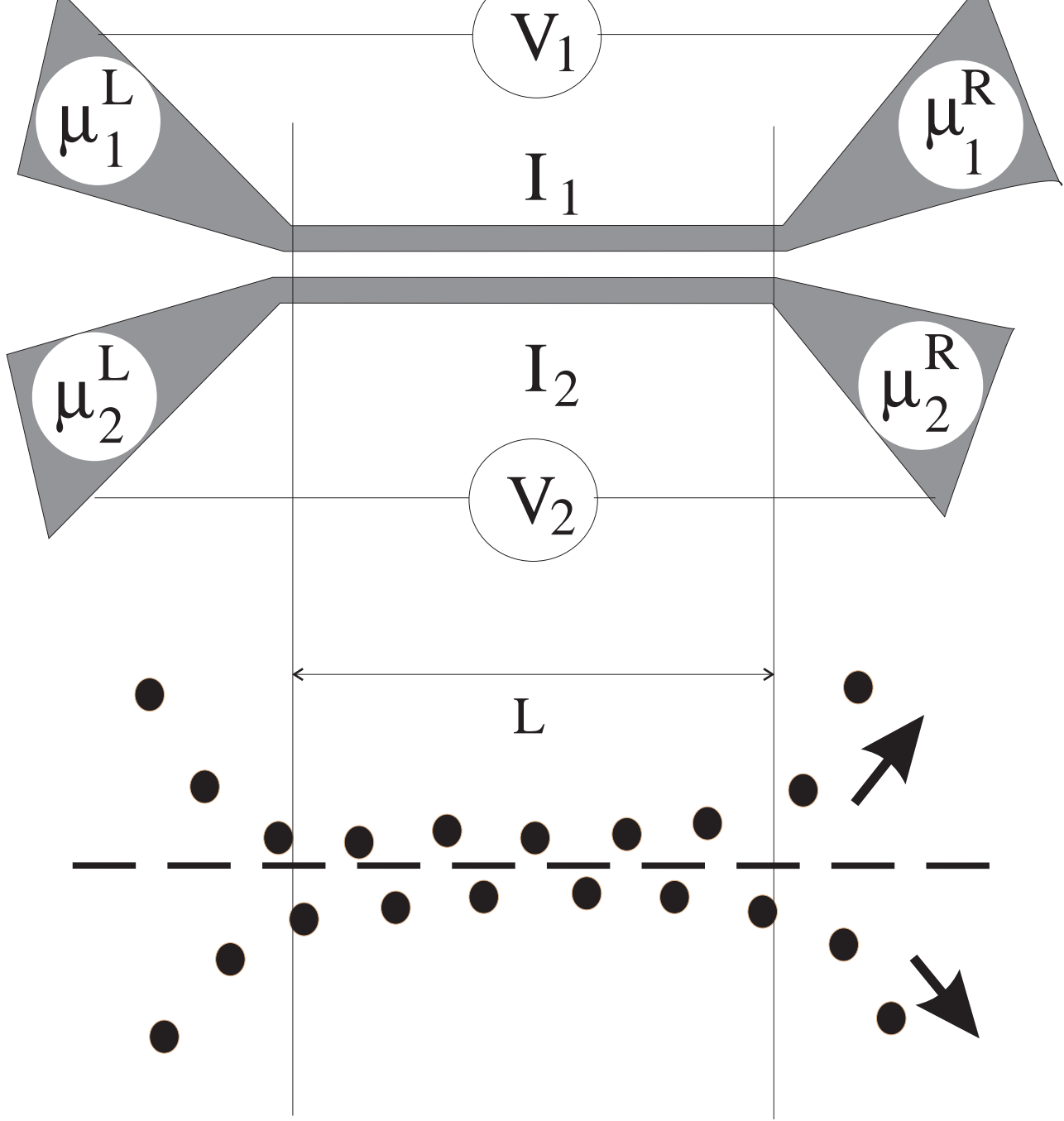
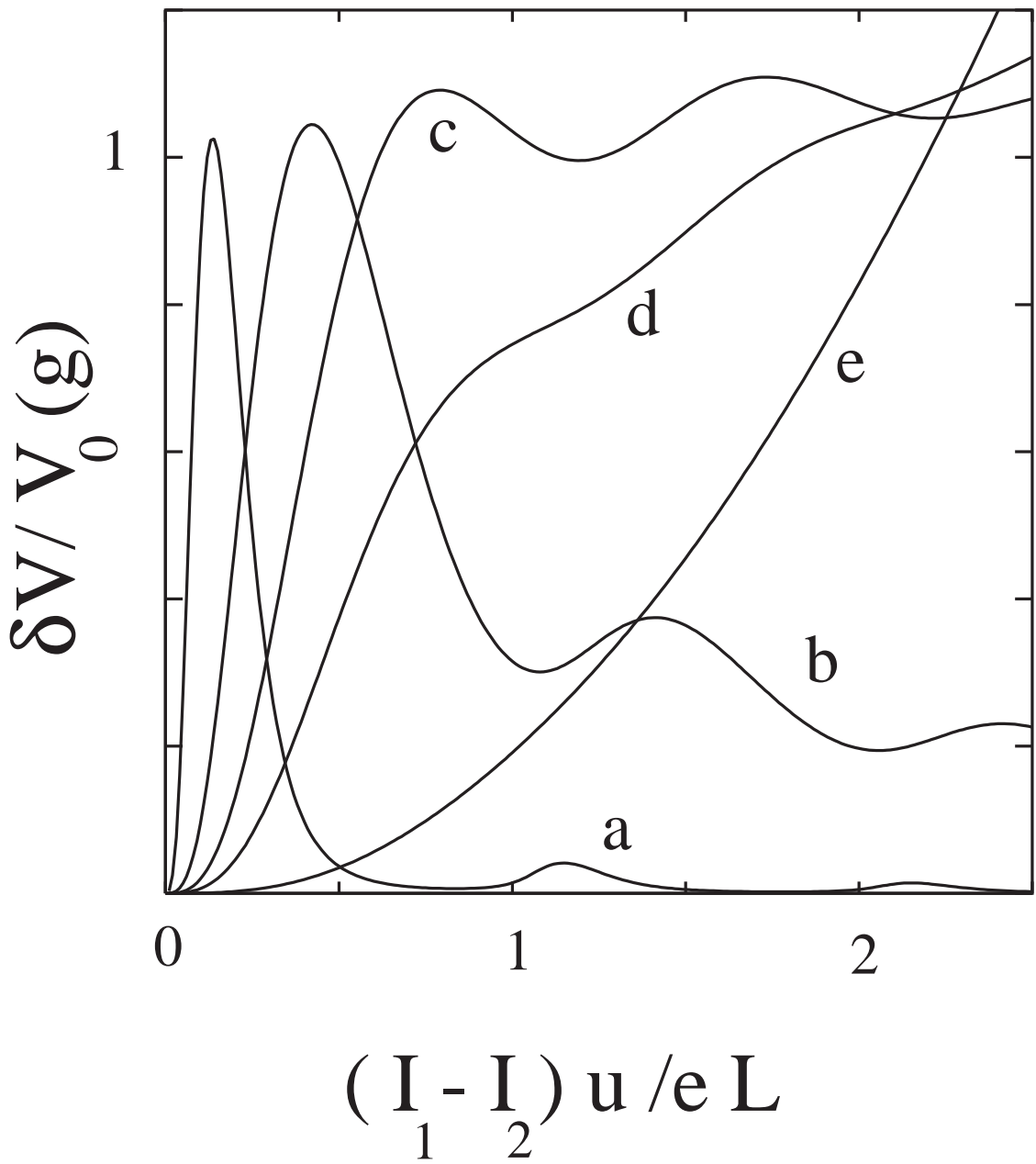
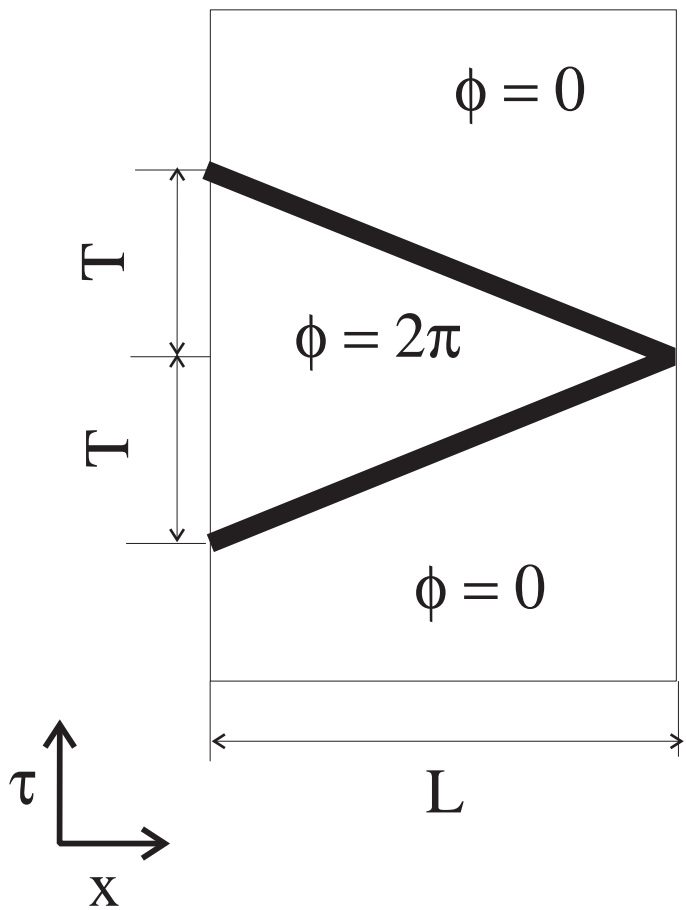
\end{document}